\documentclass[prl, twocolumn, groupedaddress,superscriptaddress]{revtex4-2}
\usepackage{graphicx}
\usepackage{amsmath}
\usepackage{amssymb}
\usepackage{amsfonts}
\usepackage{eurosym}
\usepackage{float}

\usepackage[utf8]{inputenc}
\usepackage[T1]{fontenc}
\usepackage{grffile}
\usepackage{longtable}
\usepackage{wrapfig}
\usepackage{rotating}
\usepackage[normalem]{ulem}
\usepackage{textcomp}
\usepackage{capt-of}
\usepackage{xcolor}
\usepackage{titlesec}  
\titlelabel{\thetitle.\quad}
\usepackage[
plainpages=false,                   
pdfpagelabels,                      
bookmarks=true,
bookmarksnumbered=true,
breaklinks=true,
linkbordercolor={0 0 1}]           
{hyperref} 
\usepackage{physics}
\hypersetup{hyperindex=true, colorlinks=true, urlcolor= blue, citecolor=blue, linkcolor= blue}
\usepackage[pdftex]{thumbpdf} 
\hypersetup{
 pdfauthor={Giorgio Levy},
 pdftitle={U2J in HTSC},
 pdfkeywords={},
 pdfsubject={},
 pdflang={English}}

\DeclareUnicodeCharacter{2212}{-}

\newcommand{\Tc}{$T_\text{C}$}
\newcommand{\Tcmax}{$T_\text{C}^{max}$}
\newcommand{\BE}{$BE$=-}

\begin{document}
\title{Experimental determination of superexchange energy from two-hole spectra}

\author{Giorgio Levy}
\email[E-mail: ]{levyg@phas.ubc.ca}
\affiliation{Quantum  Matter  Institute,  University  of  British Columbia,  Vancouver,  British Columbia  V6T  1Z4,  Canada}
\affiliation{Department of Physics and Astronomy, University of British Columbia, Vancouver, British Columbia V6T 1Z1, Canada}

\author{Maayan Yaari}
\affiliation{Department of Physics, Technion - Israel Institute of Technology, Haifa, 3200003, Israel}

\author{Tom Z. Regier}
\affiliation{Canadian Light Source Inc., 44 Innovation Boulevard, Saskatoon, SK, S7N 2V3, Canada}

\author{Amit Keren}
\affiliation{Department of Physics, Technion - Israel Institute of Technology, Haifa, 3200003, Israel}

\date{\today}

\begin{abstract}

  We follow the evolution of Copper and Oxygen two-hole excitations, in optimally doped  (Ca$_{x}$La$_{1-x}$)(Ba$_{1.75-x}$La$_{0.25+x} $)Cu$_{3}$O$_{y}$ for $x=0.1$ and $x=0.4$.
  The spectra have contributions from band states as well as a localized multiplet structure.
  From their identification, we determine the intrashell Coulomb interaction $U$ for Oxygen and Copper sites. These results allow us to estimate the atomic superexchange coupling $J$ suggesting a
  positive correlation between the maximal superconducting critical temperature $T_\text{C}^{max}$ and $J$.
  
\end{abstract}

\maketitle

Superconductivity in Copper-based materials (cuprates) emerges from a charge transfer insulator,
a state dominated by electronic correlations.
In the charge compensated compound
(Ca$_{x}$La$_{1-x}$)(Ba$_{1.75-x}$La$_{0.25+x} $)Cu$_{3}$O$_{y}$ (CL123), it
arises from an antiferromagnetic phase upon hole doping\cite{Bluschke2019}.
It then reaches a maximum critical temperature \Tc\ at a concentration of $p\!\simeq$0.145 holes per planar Cu. 
At a hole concentration $p\!\sim$0.12, superconductivity is suppressed and a charge order and pseudogap state appear.
However, the charge order and pseudogap energy scales do not seem to be related to superconductivity.
In contrast, a comparison
of superconducting (SC) and magnetic properties \cite{kanigel2002} suggests that \Tcmax\ grows with increasing
superexchange interaction $J$. This observation was supported by Resonant Inelastic X-ray Scattering (RIXS) \cite{Ellis2015}
and Angle Resolved Photoemission (ARPES) \cite{Drachuck2014} experiments done in the SC phase.
The ARPES data was interpreted using the fact that in the Hubbard model $J\!\sim$100 meV
increases with increasing hopping rate $t\!\sim$100 meV \cite{Jedrak2011}.

Nevertheless, none of these experiments were completely decisive.
Magnetic measurements were done in the non-SC part of the phase diagram and whether
their measured trend is extended out to the SC state could be questioned.
RIXS suffers from interpretation of the data in terms of $J$, although the arguments
seems to converge with \citet{Krzysztof2019}. In ARPES, there are two kinds of velocities,
below and above the kink in the energy dispersion $E(k)$, where ${\bf k}$ is in the $(\pi,\pi)$ direction.
The Fermi velocity (near zero energy) presents little variation with doping \cite{Bogdanov2000, zhou_universal_2003, Borisenko2006}
or between materials \cite{Edegger2006} in contrast to the high energy velocity  \cite{Drachuck2014}.
Which velocity should be compared with $T_\text{C}$ is
not clear. Moreover, the data depends on surface quality and is noisy when comparing different
cleaves even for samples of the same composition. Therefore, a convincing picture can emerge only
by performing a variety of different experiments \cite{Bogdanov2000} from which the key ingredients
guide the models for these materials \cite{spalek_universal_2017}.

The purpose of this study is to evaluate the evolution of the local superexchange interaction $J$ with $x$ in CL123, and
to compare it with $T^{max}_\text{C}(x)$. The structure of CL123 is almost identical to YBa$_2$Cu$_3$O$_y$ (Y123) \cite{Goldschmidt1993}, but
it is tetragonal with disordered chain layers. The oxygen content $y$ controls the number of doped holes,
only slightly affecting the lattice parameters \cite{Ofer2006}. Doping ranges from magnetic undoped parent
compounds to overdoped for all values of $x$. In contrast, Ca/Ba content $x$ changes only the structural
parameters such as bond lengths $a$, and Cu-O-Cu buckling angles $\theta $, while keeping the net valence
fixed \cite{sanna_experimental_2009}. The larger the $x$, the straighter and tighter is the bond.
Disorder in CL123 was found to be $x$-independent
based on the line-widths measured by: high resolution powder x-ray diffraction \cite{agrestini_soft_2014}, Cu,
Ca, and O nuclear magnetic resonance \cite{Keren_2009,Amit2010,Cvitanic2014}, phonon \cite{Wulferding2014},
and ARPES \cite{Drachuck2014}. Therefore, disorder is not responsible for variations in $T_\text{C}^{max}$.

Here, we focus on optimally doped single crystals of CL123 from two families: $x$=0.1 and $x$=0.4 where
the superconducting transition occurs at 63 and 77 K, respectively. These are the highest critical
temperatures achieved for the different $x$ values in a single crystal form.
An analysis of the Cu--$L_{3}$ and O--$K$ absorption spectra for compositions
$x$=0.1 and $x$=0.4 nearly optimally doping shows that both samples have the
same amount of holes, which is consistent with previous reports \cite{sanna_experimental_2009, agrestini_soft_2014}.
Then, by combining the previous analysis of the
absorption with the electron emission spectra, we determine the energy levels for Cu--$d$ and O--$p$ shells.
These values are then used to estimate the superexchange coupling $J$.

The measurements were performed at the Spherical Grating Monochromator (SGM) beamline located
in the Canadian Light Source (CLS) in which the samples were cleaved in a Ultra High Vacuum (UHV) environment better than $5\times 10^{-9}$ mbar.
To enhance the Auger signal we measure the spectra at the maximum of the absorption line. 
The X-Ray absorption spectroscopy (XAS) technique was performed
in the Total Electron Yield (TEY) mode, where the current needed to compensate for emitted electrons is acquired.
The incoming light beam was linearly polarized parallel to the scattering plane
defined by the incoming beam and the outgoing electrons. These were
detected using a Hemispherical Scienta analyzer. The samples were
oriented with the [001] reciprocal lattice vector perpendicular to the
analyzer entrance (in a normal emission configuration) which was also
parallel to the normal of the {\it in situ} cleaved sample surface.
\begin{figure}[h!tb]
  \includegraphics{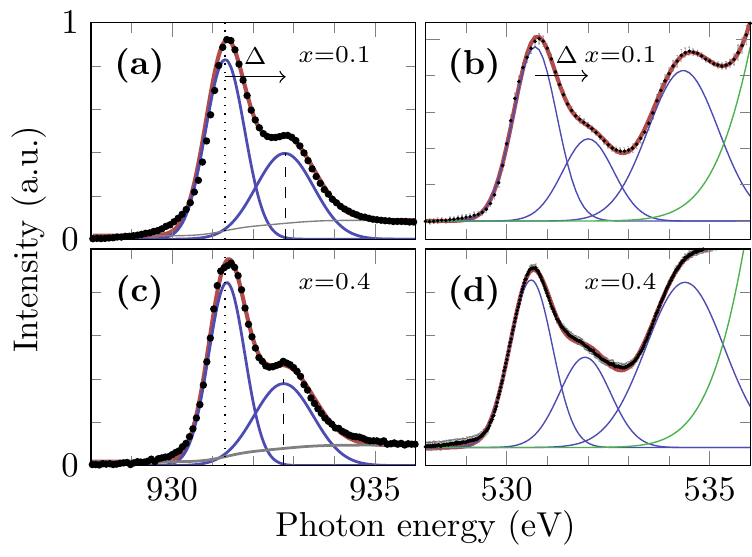}
  \centering
  \caption{{\bf Total electron yield as a function of photon energy in CL123} is plotted as black dots.
    The red curves shows the best fit to the data with a model including Gaussian profiles (blue curves) and a Shirley background
    plus an offset (black lines).
    The energy axis for Cu L$_3$ for $x$=0.1 [({\bf a})] and $x$=0.4 [({\bf c})]
    compositions where shifted to match the maximum of the spectra near 931.3eV, and the intensity of
    the main line was normalized to unity. The O K edge for $x$=0.1 is shown in  ({\bf b}) and for $x$=0.4 in  ({\bf d}).
    The arrows indicate the extracted charge transfer gap value $\Delta$.
  }
  \label{fig:L3}
\end{figure}

Prior to photoemission spectroscopy, we study the doping concentration by x-ray absorption.
The absorption spectra of the  the spin-orbit split states
near the Cu L$_3$ edge are composed by two lines: a main line at 931.3 eV and a satellite at 932.7 eV\cite{sanna_experimental_2009, agrestini_soft_2014}.
The main is assigned to the transition where a Cu--$2p$ electron is
excited to the valence band leaving a core-hole behind.
The satellite originates from this transition in the
presence of a ligand hole $\underline{L}$ in a molecular orbital formed
by the Oxygen atoms surrounding the Copper atom. The
main transitions can be represented by
$\ket{3d^9}+h\nu \rightarrow \ket{\underline{2p}, 3d^{10}}$, and the satellite by 
$\ket{3d^9,\underline{L}} +h\nu \rightarrow \ket{\underline{2p}, 3d^{10},
\underline{L}}$.

Based on the intensity of the main line and the satellite, the hole
concentration can be estimated \cite{agrestini_soft_2014, merrien_symmetry_1994}.
In poly-crystalline samples of CL123, the main line
barely varies with Calcium content $x$ or Oxygen concentration $y$\cite{agrestini_soft_2014}.
The absorption spectra acquired at the Cu $L_3$ edge are shown as black dots in Figs. \ref{fig:L3}{\bf (a)}
and {\bf (c)} for $x\!=\!0.1$ and $x\!=\!0.4$ concentrations. The corresponding ones at the Oxygen $K$ edge
are shown in Figs. \ref{fig:L3}{\bf (b)} and {\bf (d)}, respectively.
Following the analysis of \citet{agrestini_soft_2014}, we model the line-shape of the Cu L$_3$ edge with two Gaussian
profiles (blue lines in Fig. \ref{fig:L3}), a Shirley background (grey line), and an offset. Using the intensities
extracted from the fits (red lines in Fig. \ref{fig:L3}), we can estimate the total amount of
holes based on the formula: $n_h^{L_3}=I_S/(I_M+I_S)$\cite{merrien_symmetry_1994},
where $I_S$($I_M$) represents the intensity of the satellite(main)
line. We obtain a similar amount of holes \(n_h^{L_3}\!=\!0.42\!\pm\! 0.02\)  within
the error bars for both $x\!=\!0.1$ and $x\!=\!0.4$ compounds in the optimally doped regime indicating a similar
hole concentration.

The doping concentration could also be estimated from an analysis of the Oxygen $K$ (O--$K$) absorption edge.
In the similar compound Y123 \cite{nucker_site-specific_1995, hawthorn_resonant_2011}, 
this absorption edge consists of a weak pre-edge peak, associated to the O--$2p$ holes on
CuO chains followed by a structure at higher photon energies related to the Zhang-Rice state.
The following feature at increasing energies has a contribution from final states in the Upper Hubbard Band (UHB).
A similar structure is observed within $h\nu$=528--533 eV in CL123 \cite{agrestini_soft_2014}, [Fig. \ref{fig:L3} {\bf (b)} and {\bf (d)}] for both compositions $x\!=\!0.1$ and $x$=0.4.
In the chosen experimental configuration, the chains barely contribute to the spectra.
We therefore relate the feature at 530.7\textpm{0.1} eV with a transition involving a ligand hole in the CuO$_2$ planes and
the corresponding one at 532\textpm{0.1} eV with the UHB
\cite{nucker_site-specific_1995, merz_site-specific_1998, hawthorn_resonant_2011}.
The hole concentration can also be estimated from the intensity ratio of these absorption structures.
By fitting the absorption spectra with a model containing three Gaussian line-shapes
[blue lines in Fig.\ref{fig:L3}{\bf (b)} and {\bf (d)}] plus a
high-energy background, the intensity ratio
$I_{\underline{L}} /(I_{\underline{L}}+I_{UHB})$ is determined to be 0.60\textpm{0.01}, equal within error bars for both compositions.

The charge transfer gap $\Delta$ --defined as the energy separation between $3d^9$ and $3d^{10} \underline{L}$ states-- can be estimated from
the energy difference between the satellite and main line of the Cu--$L_3$ absorption edge [arrow in Fig. \ref{fig:L3}{\bf(a)}]. 
From the fits to the Cu $L_3$ spectra discussed before [Fig. \ref{fig:L3}{\bf (a)}, {\bf (c)}], we obtain for the $x\!=\!0.1$ composition a value of
$\Delta_{\text{Cu}}^{x01}=1.6\!\pm\! 0.1$ eV and for $x\!=\!0.4$,
$\Delta_{\text{Cu}}^{x04}=1.5\!\pm\!0.1$ eV.
The double peak structure appearing in the O--$K$ absorption spectra can also be interpreted as transitions to the charge transfer band at low energies, predominantly of O--$2p$ character--- followed by transitions to the UHB dominated by Cu--$3d$ character \cite{chen1991}. The energy separation between these [arrow in Fig. \ref{fig:L3}{\bf(b)}] is obtained from the fits to the absorption spectra,
$\Delta_{\text{O}}^{x01;x04}=$1.3\textpm{0.1} eV.

\begin{figure}[h!tb]
  \includegraphics{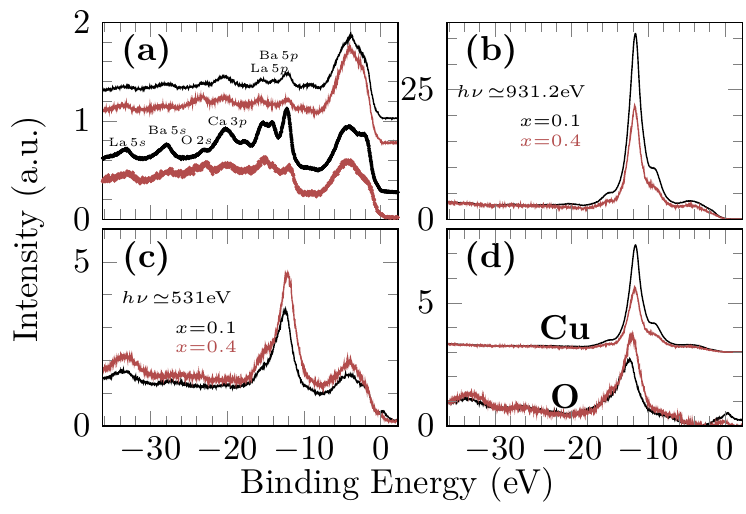}
  \centering
  \caption{ {\bf Photoemission spectra (PES) } taken around the Cu $L_3$ and O $K$ absorption edges for compositions $x$=0.1 (in black) and $x$=0.4 (in red). In panel {\bf (a)}, the spectra acquired before the absorption edges for Cu $L_3$ at $h\nu=924$ eV (thick lines) and for O $K$ at $h\nu=528$ eV (thin lines) are shifted vertically for clarity.
    Close to the absorption edges of Cu $L_3$ [{\bf (b)}, $h\nu$=931 eV]  and O $K$ [{\bf (c)}, $h\nu$=531 eV],
    the electron emission intensity has normal photoemission and Auger contributions.
    To extract the Auger component, the PES signal before the absorption edges in {\bf (a)} is subtracted to the spectra close to the maximum of absorption. The resulted spectra are assigned to the Cu $L_3M_{4,5}M_{4,5}$ [curves shifted vertically upwards in {\bf (d)}] and to O $KL_{3}L_{3}$ [bottom curves in {\bf(d)}] transitions.
     All spectra shown in this Figure have the same intensity normalization.
}
  \label{fig:res}
\end{figure}

The information extracted so far was obtained by analyzing the internal transitions during the absorption process.
To study the electronic correlations,
we concentrate on the electron emission spectra that follows the photon absorption.
By using the NIST \citet{nist_xps}, we identify the features observed in the photoemission spectra (PES) at photon energies before the absorption edges [see Fig. \ref{fig:res}{\bf (a)}]: the one at a binding energy \BE 33 eV with photoemission from La--$5s$ states,
at \BE 27.8 eV with Ba--$5s$, \BE 23.5 eV with O--$2s$, \BE 20 eV with Ca--$3p$, \BE 13.5 eV with La--$5p$, and \BE 12.8 eV with Ba--$5p$.
The difference of these features between the $x$=0.1 and $x$=0.4 composition are related to an increase of Ba to Ca concentration with $x$;
in particular, the intensity decrease at \BE 27.8 eV, as well as the increase around \BE 20 eV [Fig. \ref{fig:res}{\bf (a)}].

The PES structure in the region $BE=0-8$ eV is associated with Oxygen and Copper states whose main spectral features are centred
around -2 and -4 eV. This assignment is consistent with previous reports on HgBa$_2$Ca$_2$Cu$_3$O$_{8+\delta}$ \cite{chainani_evidence_2017} and 
Y123 \cite{Balzarotti1988}, as well as electronic structure calculations \cite{pickett_rmp}.
To better determine their binding energy,  we have fit the PES between $BE$=1 eV to $BE$=-8 eV using a model with two Gaussian line-shapes multiplied by a Fermi-Dirac function plus a Shirley background (see Supplemental Materials).
From an average of the values obtained with beam energies of $h\nu$=528 and $h\nu$=924 eV, these states are centred at \BE 1.92\textpm{0.04} eV and \BE 3.94\textpm{0.07} for the $x$=0.1 composition.
For the $x$=0.4 composition, the states are centred at \BE 1.8\textpm{0.1} and \BE 3.92\textpm{0.04} eV.
From the intensity increase with photon energy of the feature at $\sim$-2eV relative to the one at $\sim$-4eV\cite{chainani_evidence_2017}, as well as their relation with the features in the Auger spectra discussed later, we link the feature at \BE 3.9\textpm{0.1} eV with O--$2p$ partial Density of States (pDoS) and the corresponding one at \BE 1.9\textpm{0.1} eV with Cu--$3d$ pDoS. 

The electron emission spectra acquired close to the maximum of an absorption edge will include --besides the normal PES-- contributions
from different decay channels of the excited core-hole. The interference between these channels will mostly be observed in the photon energy evolution of the emission spectra \cite{gelmukhanov_resonant_1999, levy2012} where the excited core-hole has a negligible coupling with the vacuum continuum states.
As the photon energy is increased, the wave-function of the excited core-hole starts to overlap significantly with the free-propagating vacuum wave-function, which opens up the emission to Auger electrons. The emitted Auger electrons have a constant kinetic energy independent of the photon energy.
In this regime, we can approximate the electron emission as the superposition of the PES and Auger spectra \cite{kraus2013,chainani_evidence_2017}.

\begin{figure*}[h!tb]
  \includegraphics{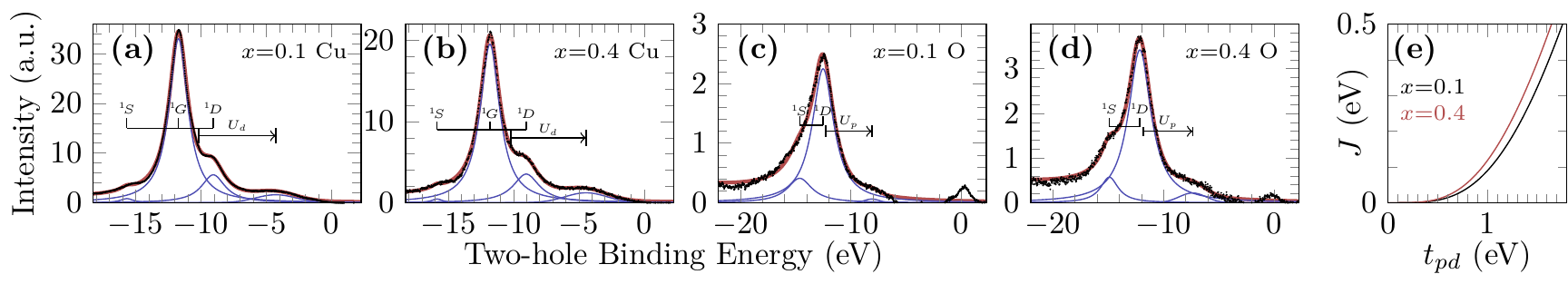}
  \centering
  \caption{ {\bf The two-hole spectra} is composed of a broad feature representing a continuum followed by a multiplet structure at higher energies. To extract these different components, we have fit the data (black dots) with a model including a Gaussian line-shape and three Lorentzian ones for the resonance close to Cu L$_3$ edge on $x$=0.1 {\bf (a)} and $x$=0.4 {\bf (b)}. For the ones close to O$K$ edge, two Lorentzian components are only included for $x$=0.1 {\bf (c)} and $x$=0.4 {\bf (d)}. Each individual line-shape is shown as a blue curve, and the final model including all the components as a red curve.
   The energy for the assigned multiplets ($^1S$, $^1D$, and $^1G$) and the obtained values for the Coulomb repulsion $U$ (arrows) are shown.
   From the on-site Coulomb repulsion as well as the charge tranfer gap, the variation {\bf (e)} of the superexchange interaction $J$
   with the charge transfer hopping $t_{pd}$ is obtained for $x$=0.1 (black line) and $x$=0.4 (red line).
    }
  \label{fig:resana}
\end{figure*}

Within 200 meV of the Cu--$L_3$ absorption threshold [Fig. \ref{fig:res}{\bf(b)}], the Cu $L_3VV$ Auger transition dominates over the
electron photoemission.
The acquired spectra is originated from the decay of the core-hole involving Cu--$d^{10}$ excited state into a final Cu--$d^8$ configuration
plus the emission of Auger electrons ($\epsilon_A$).
It consists of a broad feature centred around 4 eV plus a triple line structure around $12$ eV.
The same structure is observed for both $x$=0.1 and $x$=0.4 compositions, albeit a lower relative intensity for the last one.
When the beam energy is within 200 meV of the O--$K$ absorption edge [Fig. \ref{fig:res}{\bf(c)}], the photoemission intensity is similar to
the Auger component. Notwithstanding, a clear double line contribution around 12 and 15 eV can be observed on the Auger spectra.
This component can be extracted by removing the PES acquired at photon energies lower than the absorption edge.
This is performed based on the photon energy evolution of the acquired electron emission vs. binding energy spectra \cite{kraus2013}.
Then, the Auger spectra originated from the core-hole decay involving excited O--$2p^6$ states into a O--$2p^4$ final configuration can be clearly observed. It consists of a broad feature around 7 eV plus a double line structure [bottom curves on Fig. \ref{fig:res}{\bf (d)}]
which becomes evident for both $x$=0.1 and $x$=0.4 compositions. 
The spectra close to the O--$K$ absorption edge also present a structure around $BE\!\simeq$0 eV which originate from the attenuated higher harmonics from the synchrotron beam.
When the same extraction procedure is applied to the spectra acquired  close to the Cu--$L_3$ edge [top curves on Fig. \ref{fig:res}{\bf(d)}], the broad feature around $4$ eV is mostly highlighted.
A first inspection at the position of the spectroscopic lines shows that the maximum of the Auger spectra for Cu $L_3VV$ transition is at a lower  energy than the corresponding O $KVV$ one.
The spectroscopic signatures of the Auger structure are then extracted by fitting the spectra shown in Fig. \ref{fig:res}{\bf(d)} where
the broad features are modelled by a Gaussian line-shape, and the multiple lines by Lorentzian ones [see Fig. \ref{fig:resana}].

The extracted spectroscopic information is interpreted using Cini-Sawatzky theory (CST) \cite{cini_two_1977, sawatzky1977} as a framework.
In this theory, the Auger emission from an atomic closed shell embedded in
a solid consists of bound states separated from band states by the intrashell Coulomb repulsion $U$.
Therefore, in the two--hole spectra, the broad feature is assigned to transitions from band states, and the relatively narrower lines
to emission from bound state multiplets.
Furthermore, because the direct transition between a spin singlet and triplet configuration of two electrons is forbidden for a spherically
symmetric operator -- propensity rules \cite{agren_spin-orbit_1993, kyiene_maximal_2004}-- the lines are assigned to final states in
the spin-singlet configuration only. For Cu, these would correspond to $^1G$, $^1D$, and $^1S$ multiplets and for O, to $^1D$ and $^1S$. 
Finally,  owing to their spin-singlet configuration, the total spin-orbit coupling for these final states is quenched.

Based on these considerations, we can extract the effective intrashell Coulomb interaction for Cu and O elements from the two-hole spectra.
First, the centre and width of the band states are extracted from a fit to the broad features.
This assignment is further supported by linking these values with the corresponding pDoS obtained before; the band states are theoretically described by the self-convolution density of states (SCDoS) of the respective pDoS. 
For $x$=0.1 composition, the band states related to Cu are centred at 4.27\textpm{0.03} eV with a full width at half maximum (FWHM) of 3.64\textpm{0.06} eV.
Second, by associating the lines at increasing two-hole energy with the multiplets  $E(^1D)$=9.06\textpm{0.01} eV, $E(^1G)$=11.72\textpm{0.01} eV, and $E(^1S)$=15.67\textpm{0.03} eV, we determine the Slater-Condon parameters \cite{griffith1961} (see Supplemental Materials). From them, the effective
intrashell Coulomb interaction $U$ is extracted as the difference between the Slater-Condon parameter $F^0$ and the centre of the band states \cite{sawatzky1980}; which results in $U^{x01}_d\!\simeq$5.91\textpm{0.03} eV for $x$=0.1.
For $x$=0.4, the band states are centred at 4.45\textpm{0.04} eV with a FWHM of 3.8\textpm{0.1} eV. From this value and the position of the multiplets, $E(^1D)$=9.03\textpm{0.01} eV,
 $E(^1G)$=11.81\textpm{0.01} eV,
and $E(^1S)$= 15.86\textpm{0.03} eV, we obtain $U^{x04}_d\!\simeq$5.76\textpm{0.04} eV.
In O $KVV$ spectra, the multiplets  are located at $E(^1D)$=12.52\textpm{0.01} eV and $E(^1S)$=14.64\textpm{0.05} eV for
$x$=0.1 [Fig. \ref{fig:resana}{\bf(c)}] composition. An effective O--$2p$ intrashell Coulomb repulsion $U_p^{x01}$=4.18\textpm{0.02} eV
relative to the band states is obtained. 
For $x=0.4$ composition, the multiplets are located at $E(^1D)$=12.14\textpm{0.01} and $E(^1S)$=14.90\textpm{0.03} eV giving $U_p^{x04}$=4.47\textpm{0.06} eV. 
Even though the obtained intrashell Coulomb repulsion for Cu--$3d$ states is smaller than the reported values for Bi$_2$CaCu$_2$O$_8$ \cite{tjen1992} ($U_d\sim$8 eV), for La$_{1.85}$Sr$_{0.15}$CuO$_{4}$ and La$_{1.85}$Ba$_{0.15}$CuO$_{4}$ \cite{barderoma1992} ($U_d\!\simeq$7--8 eV), it is within the range for the reported value in Y123 \cite{marel1988} ($U_d\!\simeq$5--7eV). Regarding the intrashell Coulomb interaction for O--$2p$ states, our values are similar to the previously reported $U_p\!\simeq$5 eV \cite{tjen1992, barderoma1992, marel1988} in different compounds. This indicates the existence of strong electronic correlations in CL123 similar to other copper-based superconductors.

The atomic antiferromagnetic superexchange coupling $J$ is revealed as one manifestation of the electronic correlations. This coupling would mediate the spin-spin interaction between two adjacent Cu atoms through an intermediate O. Based on a simple atomic transition model \cite{pavarini_book} which neglects the details of the band-structure \cite{eskes1993}, we numerically determine the superexchange energy $J$; which depends on the Coulomb repulsion between two electrons in O--$2p$ orbitals ($U_p$), the corresponding one in Cu--$3d$ orbital ($U_d$), the onsite energy difference $\Delta_\text{pd}$ between them, and the charge transfer hopping $t_{pd}$ (see Supplemental Materials).
The trend of $J$ with $t_{pd}$ [Fig. \ref{fig:resana}{\bf(e)}], where the other parameters are extracted from the Cu absorption and the two-hole spectra, indicates a higher value for $x$=0.4 than for $x$=0.1 composition. At fixed value of $t_{pd}$, the increase of the superexchange coupling $J$ is  driven by the variation of the Coulomb interaction with $x$: an increase of the intrashell correlations in O--$2p$ orbitals and a decrease of the ones in Cu--$3d$.
On the other hand, $t_{pd}$ can be estimated from the previously determined values of $J$\cite{Ofer2006}:
for $x$=0.1, $J$=82\textpm{5}meV and for $x$=0.4, $J$=115\textpm{7} meV; which results in $t_{pd}$=0.98\textpm{0.02} eV and 1.00\textpm{0.02} eV, respectively.
Thus $J$ increases with $x$ owing  mainly to the electronic correlation variations.

In summary, guided by CST and the propensity rules for Auger transitions, we have identified the band states as well as spin-singlet multiplet excitations of the extracted two-hole spectra of Copper and Oxygen elements on the high-$T_\text{C}$ compound (Ca$_{x}$La$_{1-x}$)(Ba$_{1.75-x}$La$_{0.25+x} $)Cu$_{3}$O$_{y}$. Based on their energy, the Coulomb repulsion for electrons in the O--$2p$ ($U_p$) and
Cu--$3d$ ($U_d$) states was determined for the compositions $x$=0.1 and $x$=0.4 in the optimally doped regime. Using these values together with the charge-transfer gap extracted from the X-Ray absorption spectra, we have indicated that the atomic superexchange interaction $J$ increases with $x$ as does $T_\text{C}^{max}$. This relation is consistent with the recently reported \cite{wang_paramagnons_2020} connection between $T_\text{C}^{max}$ and $J$ for different high-$T_\text{C}$ families.

{\bf Acknowledgments:} We thank J. Spa\l{}ek, A. Nocera, and D. Ellis for useful discussions.
This work was performed at the Canadian Light Source, Saskatchewan, Canada.
This study was also
supported by the Canada First Research Excellence Fund,
Quantum Materials and Future Technologies Program; and the Israeli Science Foundation individual grant
program 315/17.

\bibliographystyle{apsrev4-2}
%

\end{document}